\documentclass[11pt,a4paper]{article}
\usepackage[margin=25mm]{geometry}
\usepackage[utf8]{inputenc}
\usepackage[T1]{fontenc}
\usepackage{lmodern}
\usepackage{amsmath,amssymb,bm}
\usepackage{graphicx}
\usepackage{booktabs}
\usepackage{caption}
\usepackage{subcaption}
\usepackage{natbib}
\usepackage{setspace}
\usepackage{lineno}
\usepackage[hidelinks]{hyperref}
\usepackage{microtype}
\usepackage{xcolor}
\usepackage{adjustbox}
\graphicspath{{figures/}}
\DeclareMathOperator*{\argmax}{arg\,max}
\DeclareMathOperator{\median}{median}
\DeclareMathOperator{\logit}{logit}
\DeclareMathOperator{\Gap}{Gap}
\newcommand{\bS}{\mathbf{S}}
\title{Clustering trajectories of modelled estimates: an uncertainty-aware framework, with an application to global cardiovascular disease burden, 1990 to 2023}
\author{Eben Afrifa-Yamoah\thanks{ Mathematical Applications and Data Analytics, School of Science, Edith Cowan University, Perth, Western Australia. Correspondence: e.afrifayamoah@ecu.edu.au. ORCID 0000-0003-1741-9249.}}
\date{}

\begin{document}
\maketitle

\begin{abstract}
\noindent Modelled health estimates such as those of the Global Burden of Disease (GBD) study are increasingly used in secondary analyses designed for observed data. However, trajectory clustering typically ignores published uncertainty, selects the number of clusters using criteria blind to it, and assigns countries to groups with unwarranted certainty. We propose an uncertainty-aware framework that separates trajectory level from shape, quantifies membership probabilities by perturbing estimates within their uncertainty intervals, selects clusters based on stability under uncertainty, and examines the roles of development, region, and measurement quality. Applied to age-standardised cardiovascular disease DALY rates for 201 countries from 1990-2023 (GBD 2023), the framework identified two stable trajectory types, whereas the gap statistic selected three. The additional cluster was poorly supported (median membership probability 0.70) and disappeared once uncertainty was considered. The retained clusters had similar burdens in 1990 but diverged markedly over time: a stalled-improvement group of 78 countries (median change $-8\%$, 40 of them in sub-Saharan Africa) and a sustained-decline group of 92 countries ($-52\%$). Thirty-one countries remained unclassified because their memberships were uncertain. Results were reproduced using mortality rates, a 2000-2023 analysis window, and bootstrap resampling. Among countries with the narrowest uncertainty intervals, a third well-supported trajectory emerged in former Soviet and Eastern European countries. These findings show that uncertainty affects both the number and composition of trajectory clusters. Ignoring it risks over-interpreting unstable groups and overstating confidence in country classifications. Incorporating published uncertainty and measurement quality should become standard practice when clustering modelled health estimates. 

\medskip\noindent\textbf{Keywords:} trajectory clustering; uncertainty propagation; cluster stability; Global Burden of Disease; cardiovascular disease; modelled estimates
\end{abstract}

\section{Introduction} 
Modelled health estimates are increasingly used as data in secondary analyses. The Global Burden of Disease (GBD) study alone provides annual estimates of mortality, prevalence, incidence and disability for more than 200 countries and territories, creating unprecedented opportunities to study long-term population trajectories \citep{gbd2023dalys,gbd2023cod}. However, these estimates are not observations. They are outputs of a statistical modelling framework that synthesises vital registration, verbal autopsy, surveys and covariates, with the balance of evidence varying substantially across countries. Every estimate is accompanied by a 95\% uncertainty interval derived from 1,000 posterior draws, and those intervals are widest where primary data are weakest. Under the GATHER guidelines, uncertainty is intended to accompany all reported estimates \citep{stevens2016}. Recent commentaries have nevertheless questioned the growing practice of conducting secondary analyses on GBD point estimates alone, often without considering whether uncertainty is large enough to affect the resulting conclusions \citep{munozlaguna2024,husoy2025}. This paper addresses that concern for one of the most common forms of secondary analysis: trajectory clustering. 

Existing clustering analyses of GBD trajectories almost invariably treat the published point estimates as the data. A recent global study clustered joint HIV and tuberculosis mortality trajectories for 204 countries using point estimates alone \citep{mobaderi2025}, and the same practice is evident across a growing literature applying trajectory methods to GBD series. The clustering methods themselves are well developed. Comparative evaluations have established the circumstances under which growth mixture models, two-step growth curve clustering and longitudinal k-means recover known trajectory groups \citep{denteuling2023}, and modern software makes these approaches readily interchangeable \citep{denteuling2025}. What these methods generally do not incorporate is uncertainty in the observed trajectories. As a result, the number of clusters is typically chosen by geometric criteria such as the silhouette statistic \citep{rousseeuw1987} or gap statistic \citep{tibshirani2001}, both of which are blind to uncertainty; all countries are assigned with equal confidence; and clusters supported largely by heavily modelled estimates cannot easily be distinguished from those supported by strong underlying data. 

The statistical literature has long recognised the importance of cluster stability. \citet{hennig2007} proposed cluster-wise stability assessments based on resampling units, an idea now widely used for evaluating clustering solutions and selecting the number of clusters \citep{fang2012,denteuling2025}. Simulation studies further show that measurement error can create, obscure or distort cluster structure depending on the clustering algorithm employed \citep{pankowska2025}. Measurement error has also been incorporated directly into clustering procedures. For example, \citet{kumar2007} modify Ward's method and k-means by weighting distances according to measurement-error covariance, reducing the influence of uncertain observations on the final partition. Our objective is different. Rather than altering the clustering algorithm itself, we ask how strongly the published uncertainty attached to each trajectory affects confidence in cluster membership and in the selected number of clusters. Where traditional stability analyses resample units, we resample values within their published uncertainty intervals. This distinction allows the stability assessment to reflect the uncertainty structure of the estimates themselves. 

We develop an uncertainty-aware framework for clustering trajectories of modelled estimates. The framework separates trajectory level from trajectory shape so that clusters reflect temporal patterns rather than absolute burden; quantifies membership probabilities through repeated perturbation and reclustering of trajectories within their uncertainty intervals; selects the number of clusters according to stability under uncertainty; and evaluates the contribution of development level, region and measurement quality to the resulting classifications. The framework is designed to be general and applicable wherever published estimates are accompanied by uncertainty intervals. 

We demonstrate the framework using age-standardised cardiovascular disease burden from the Global Burden of Disease study. Cardiovascular disease remains the leading cause of disease burden worldwide, accounting for 19.2 million deaths and 437 million DALYs in 2023, with approximately a sixteenfold difference in age-standardised DALY rates between countries \citep{gbdcvd2025}. The cardiovascular transition has unfolded unevenly across the world. In many high-income countries, decades of decline in cardiovascular mortality have slowed since the early 2010s \citep{lopez2019,dowd2025}, whereas many low- and middle-income countries experienced continued increases in burden well after rates had peaked elsewhere. In Africa, for example, the age-standardised mortality rate from non-communicable diseases declined by only 9\% between 1990 and 2023 while the absolute number of deaths more than doubled \citep{gbdafrica2026}. These heterogeneous trajectories provide a substantive setting in which uncertainty-aware clustering can reveal whether apparent groups represent robust population patterns or artefacts of estimation uncertainty. 

This paper makes three contributions. First, it develops a reproducible framework for clustering trajectories of modelled estimates while explicitly propagating published uncertainty to both cluster membership and the choice of the number of clusters. Second, applying the framework to global cardiovascular disease burden, it shows that a conventional clustering criterion identifies a trajectory group that is not supported once uncertainty is taken into account, while revealing two robust trajectory types that are reproducible across burden measures, analysis windows and resampling schemes. Third, it demonstrates that measurement quality materially shapes the resulting partition: countries with wider uncertainty intervals are disproportionately represented in the stalled trajectory, while analyses restricted to well-measured countries reveal additional structure obscured in the pooled data. 

Section~\ref{sec:data} describes the data and their preparation. Section~\ref{sec:methods} develops the framework and justifies its design choices. Section~\ref{sec:results} presents the case study and robustness analyses. Section~\ref{sec:discussion} discusses implications, limitations and directions for future work.

\section{Data}\label{sec:data}

The primary series is the age-standardised rate of disability-adjusted life years (DALYs) per 100,000 for cardiovascular diseases (GBD level-2 cause), both sexes, in each of 204 countries and territories for every year from 1990 to 2023, from GBD 2023 \citep{gbd2023dalys}. DALYs combine fatal and non-fatal burden and are the measure GBD leads with; the age-standardised death rate for the same cause, years and countries was obtained for a sensitivity analysis. Age standardisation uses the GBD world standard population and is essential for comparing trajectories across countries, since a rising crude rate in an ageing population can coincide with a falling age-standardised one.

Covariates were the Socio-demographic Index (SDI) for 2023, a composite of income per capita, mean years of schooling and fertility under 25 on a zero to one scale \citep{gbd2023dalys}; the GBD super region (seven groups); and the World Bank income group used to partition the extraction. Country centroids for the spatial analysis were taken from the largest polygon of each country in the \texttt{maps} package for R.

\subsection{Access and processing}

All GBD values were retrieved programmatically through the IHME GBD 2023 Model Context Protocol connector rather than from the GBD Results Tool, and the route shaped the analysis in two ways. The connector caps each response at roughly 667 location years for DALYs and 476 for deaths in its default format, so each global series was assembled from 12 calls, one per World Bank income group and year window, and reconciled to 203 countries with 34 complete years. The connector also supplies the 95\% interval for the all-ages rate but not for the age-standardised rate, and does not expose the posterior draws; the uncertainty propagation in Section~\ref{sec:stability} therefore rests on an approximation described there, and a draw-level replication from the Results Tool export remains the first item for revision.

The analysis set is 201 countries: the 204 GBD locations less Hong Kong and Macao, which are World Bank economies but sit under China in the GBD hierarchy, and Tuvalu, which lacks an income classification in the connector and a polygon in the base map. All rates were log transformed before analysis.

\section{Methods: an uncertainty aware framework}\label{sec:methods}

The framework has four components, applied in sequence. Each is described with the decision it embodies, the alternatives we considered, and why we chose as we did. The guiding principle is that every quantity the framework reports, including the number of clusters, should have been tested against the uncertainty that the data provider published.

\subsection{Notation and inputs}

Let $i = 1, \ldots, n$ index countries and $t = 1, \ldots, T$ index years, with $n = 201$ and $T = 34$. For each country year, GBD publishes a point estimate $r_{it}$ of the age-standardised rate and a 95\% uncertainty interval $[\ell_{it}, u_{it}]$ obtained as the 2.5th and 97.5th percentiles of 1,000 posterior draws. We work throughout on the log scale, $y_{it} = \log r_{it}$, because rates are strictly positive, their uncertainty is approximately symmetric on that scale, and differences in $y$ are proportional changes that compare across countries with very different burden. The published interval is converted to a per-observation standard deviation on the log scale by treating the posterior of $r_{it}$ as log-normal,
\begin{equation}\label{eq:sd}
s_{it} = \frac{\log u_{it} - \log \ell_{it}}{2\, z_{0.975}}, \qquad z_{0.975} = 1.96.
\end{equation}
Equation~\eqref{eq:sd} recovers the posterior standard deviation exactly if the draws are log-normal and approximately otherwise. It is the only information about uncertainty the framework needs, and it is available for every GBD output. Its country mean,
\begin{equation}\label{eq:width}
w_i = \frac{1}{T}\sum_{t=1}^{T} s_{it},
\end{equation}
is used in Section~\ref{sec:attribution} as a measure of how far country $i$'s series depends on modelling rather than on primary data.

\subsection{Level and shape}

A trajectory has a magnitude and a form, and clustering on the raw series conflates them: two countries with the same decline but burden differing threefold sit far apart, while a high burden country that is stalling can sit next to one that is falling fast. We separate the two by an orthogonal decomposition of each series into a level and a shape,
\begin{equation}\label{eq:decomp}
y_{it} = L_i + S_{it}, \qquad L_i = \frac{1}{T}\sum_{t=1}^{T} y_{it}, \qquad S_{it} = y_{it} - L_i,
\end{equation}
so that $\sum_t S_{it} = 0$ for every country. The shape vector $\bS_i = (S_{i1}, \ldots, S_{iT})^\top$ is the log ratio of the rate in each year to the country's own average over the period; it is dimensionless, proportional, and invariant to any constant rescaling of the series. Clustering uses $\bS_i$ alone, and $L_i$ is kept as a diagnostic.

Two alternatives were considered. First differences, $\Delta y_{it} = y_{it} - y_{i,t-1}$, remove level and any constant trend and leave acceleration; in these data they gave a silhouette of 0.09 at $k = 2$ and no interpretable structure, because most of the signal is in the trend itself. Fitting each series with a low-order polynomial or B spline and clustering the coefficients is attractive for noisy series but adds a tuning choice and, with 34 annual points and smooth GBD trajectories, offered no reduction worth the cost. Mean centring is the simplest option that isolates form, and we check after clustering that it worked. For a partition $\mathcal{C} = \{C_1, \ldots, C_k\}$ with $c(i)$ the cluster of country $i$, the share of level variance explained by membership,
\begin{equation}\label{eq:r2level}
R^2_L = 1 - \frac{\sum_i \bigl(L_i - \bar L_{c(i)}\bigr)^2}{\sum_i \bigl(L_i - \bar L\bigr)^2},
\end{equation}
should be near zero, and the adjusted Rand index \citep{hubert1985} between $\mathcal{C}$ and the tertiles of $L$ should be near zero.

\subsection{Clustering}

Shape vectors are clustered by agglomerative hierarchical clustering with Ward's criterion on Euclidean distance $d(i,j) = \lVert \bS_i - \bS_j \rVert_2$. At each step Ward's method merges the pair of clusters $A$ and $B$ whose union least increases the total within-cluster sum of squares,
\begin{equation}\label{eq:ward}
\Delta(A, B) = \frac{|A|\,|B|}{|A| + |B|}\, \lVert \bar{\bS}_A - \bar{\bS}_B \rVert_2^2,
\end{equation}
where $\bar{\bS}_A$ is the centroid of $A$. We use the \texttt{ward.D2} implementation, which applies the criterion to squared distances as Ward intended \citep{murtagh2014}. Ward's method produces compact groups of roughly equal variance and a full dendrogram, so the partition $\mathcal{C}_k$ for every $k$ comes from one tree by cutting at $k$ branches, which simplifies the stability analysis. The partition was checked against partitioning around medoids at the same $k$; agreement was moderate (adjusted Rand index 0.72 at $k = 2$), which is expected when a share of countries sit near a boundary and is one reason to report membership probabilities rather than hard labels.

Euclidean distance on centred series treats every year equally and cannot align two countries whose transition occurs at different dates. Dynamic time warping would allow that alignment; we did not use it because the timing of a country's turning point is itself of interest and warping removes it. Two alternatives are left for future work. An error-weighted distance in the manner of \citet{kumar2007} would let the published intervals shape the partition directly rather than only its assessment; we chose not to, because the point estimate partition is what existing analyses report and the framework is meant to audit it. A codispersion based dissimilarity, which measures co-variation of two series at a chosen lag rather than pointwise distance \citep{fouedjio2026}, is a promising alternative for regionalised series.

\subsection{Membership probability and the number of clusters}\label{sec:stability}

This component makes the framework uncertainty-aware, and it rests on one observation: each $y_{it}$ comes with a published interval, so the natural perturbation is of the value, not of the sample of countries. For $b = 1, \ldots, B$ we draw a perturbed panel
\begin{equation}\label{eq:perturb}
y^{(b)}_{it} = y_{it} + s_{it}\,\varepsilon^{(b)}_{it}, \qquad \varepsilon^{(b)}_{it} \overset{\text{iid}}{\sim} \mathcal{N}(0,1),
\end{equation}
recompute level and shape by~\eqref{eq:decomp}, and recluster at the same $k$ to obtain a perturbed partition with labels $c^{(b)}(i)$. Because cluster labels are arbitrary, each perturbed partition is aligned to the reference by majority overlap: perturbed cluster $m$ is mapped to the reference cluster it shares most members with,
\begin{equation}\label{eq:align}
\sigma^{(b)}(m) = \argmax_{c \in \{1,\ldots,k\}} \bigl| C^{(b)}_m \cap C_c \bigr|.
\end{equation}
The membership probability of country $i$ is the share of draws in which it returns to its reference cluster under that alignment,
\begin{equation}\label{eq:pmember}
p_i = \frac{1}{B}\sum_{b=1}^{B} \mathbf{1}\Bigl\{ \sigma^{(b)}\bigl(c^{(b)}(i)\bigr) = c(i) \Bigr\}.
\end{equation}
We used $B = 1{,}000$. The map~\eqref{eq:align} is many to one, so two perturbed clusters can be sent to the same reference cluster; this is deliberate, since a reference cluster that splits under perturbation should register as unstable for the members that leave, and it does.

Three decisions follow from~\eqref{eq:pmember}. A country is reported as assigned only if its membership probability clears a threshold $\tau$,
\begin{equation}\label{eq:assign}
a_i = \begin{cases} c(i) & \text{if } p_i \geq \tau, \\ 0 & \text{otherwise,} \end{cases} \qquad \tau = 0.70,
\end{equation}
where $a_i = 0$ denotes unassigned; $\tau$ is in the range Hennig recommends for a stable cluster under resampling \citep{hennig2007}, and the number of unassigned countries is reported. The number of clusters is the largest $k$ for which every cluster is stable on its own terms,
\begin{equation}\label{eq:kstar}
k^{*} = \max\Bigl\{ k : \min_{c \in \{1,\ldots,k\}} \median_{i \in C_c}\, p^{(k)}_i \geq \pi \Bigr\}, \qquad \pi = 0.80,
\end{equation}
with $p^{(k)}_i$ computed under the $k$-cluster partition. This is deliberately conservative relative to the two conventional criteria reported alongside it. The silhouette \citep{rousseeuw1987} rewards separation of point estimates. The gap statistic \citep{tibshirani2001} compares the log pooled within-cluster dispersion $W_k = \sum_c \sum_{i \in C_c} \lVert \bS_i - \bar{\bS}_c \rVert_2^2$ with its expectation under a uniform reference,
\begin{equation}\label{eq:gap}
\Gap(k) = \mathbb{E}^{*}\bigl[\log W_k\bigr] - \log W_k,
\end{equation}
and selects the smallest $k$ with $\Gap(k) \geq \Gap(k+1) - \hat\sigma_{k+1}$, the expectation and its standard error coming from 100 reference sets. Neither criterion knows how far each point could move; \eqref{eq:kstar} does, and can reject a partition that \eqref{eq:gap} accepts.

The same machinery runs with units resampled instead of values. In the country bootstrap, draw $b$ takes a bootstrap sample of countries, clusters the unique members, and aligns as in~\eqref{eq:align}; $p_i$ is then the share of the draws containing $i$ in which it returns to its reference cluster. This is Hennig's scheme, and reporting both lets a reader see which source of instability dominates.

The perturbation model~\eqref{eq:perturb} is independent across years and countries. That is a simplification: posterior draws within a country are correlated across years, and a fully coherent treatment would resample from the GBD draw files. Independence produces rougher perturbed trajectories than the truth and so, if anything, understates stability; it is the conservative direction. A second approximation is specific to the access route: the age-standardised rate carries no interval in the connector output, so $s_{it}$ in~\eqref{eq:sd} was computed from the all-ages rate's interval. Both approximations are removed by replication on draw files.

\subsection{Attribution}\label{sec:attribution}

A cluster solution invites the question of what it tracks. We answer in three stages. Spatial coherence is assessed on the assigned countries $\mathcal{A} = \{i : a_i > 0\}$, $n_{\mathcal{A}} = |\mathcal{A}|$. Let $d_{ij}$ be the great-circle distance between the centroids of $i$ and $j$, and define a row-standardised weight matrix $\mathbf{W}$ with $w_{ij} = 1/\kappa$ if $j$ is among the $\kappa$ nearest neighbours of $i$ and zero otherwise. With $z_i = \mathbf{1}\{a_i = c\} - \bar z$ the centred indicator of cluster $c$, Moran's I \citep{moran1950} is
\begin{equation}\label{eq:moran}
I_c = \frac{n_{\mathcal{A}}}{\sum_i \sum_j w_{ij}} \cdot \frac{\sum_i \sum_j w_{ij}\, z_i z_j}{\sum_i z_i^2},
\end{equation}
tested against 999 random permutations of the labels, and reported for $\kappa \in \{4, 5, 6, 8, 10\}$ with $\kappa = 6$ as the primary choice, the smallest number that connects every island state to the mainland graph.

Development versus region. For the two-cluster solution we fit nested logistic models for membership of the decline cluster among assigned countries,
\begin{equation}\label{eq:logit}
\logit \Pr(a_i = 2) = \beta_0 + \beta_1\,\mathrm{SDI}_i \;[\,+ \textstyle\sum_{r=2}^{7} \gamma_r\,\mathbf{1}\{\mathrm{region}_i = r\}\,] \;[\,+ \delta\, w_i\,],
\end{equation}
adding super region and the interval width $w_i$ of~\eqref{eq:width} in turn, and report McFadden's pseudo $R^2 = 1 - \ell(\mathcal{M})/\ell(\mathcal{M}_0)$ and likelihood ratio tests for each added term. The first model asks how much of the partition is development; the second whether geography adds anything once development is held constant; the third whether measurement quality predicts membership once both are held constant, which is the test for a modelling artefact that the data allow.

The unassigned class. Because~\eqref{eq:assign} produces three outcomes, we also fit a multinomial logistic model of $a_i \in \{1, 2, 0\}$ on SDI and region over all 201 countries, and plot the predicted probability of each outcome across the SDI range. If the unassigned countries are those genuinely in transition rather than noise, the probability of being unassigned should peak where the stalled and decline probabilities cross.

\subsection{Robustness design and implementation}

Four analyses test whether the primary partition is a property of the data or of the choices made. The framework is rerun in full on age-standardised death rates in place of DALYs; on DALYs over 2000 to 2023 only, the window in which the high-income slowdown documented by \citet{lopez2019} and \citet{dowd2025} would show as a late-stalling trajectory type; under the country bootstrap in place of value perturbation; and on the half of countries with the narrowest published intervals, to see what structure the best-measured series carry on their own. Agreement with the primary partition is reported as the adjusted Rand index over countries assigned in both. The assignment threshold $\tau$ and the stability rule $\pi$ are varied over a grid.

All analysis was run in R 4.3.3 with base R and the packages \texttt{cluster}, \texttt{nnet} and \texttt{maps}. Equations~\eqref{eq:sd} to~\eqref{eq:kstar} and~\eqref{eq:moran} are implemented directly; \eqref{eq:gap} uses \texttt{clusGap}, \eqref{eq:logit} uses \texttt{glm}, and the multinomial model uses \texttt{multinom}. One seed and one run of $B = 1{,}000$ draws produced every table and figure; the run takes about 75 seconds on a laptop.

\section{Results}\label{sec:results}

\subsection{Number of clusters}

\begin{figure}[t]
\centering\includegraphics[width=\textwidth]{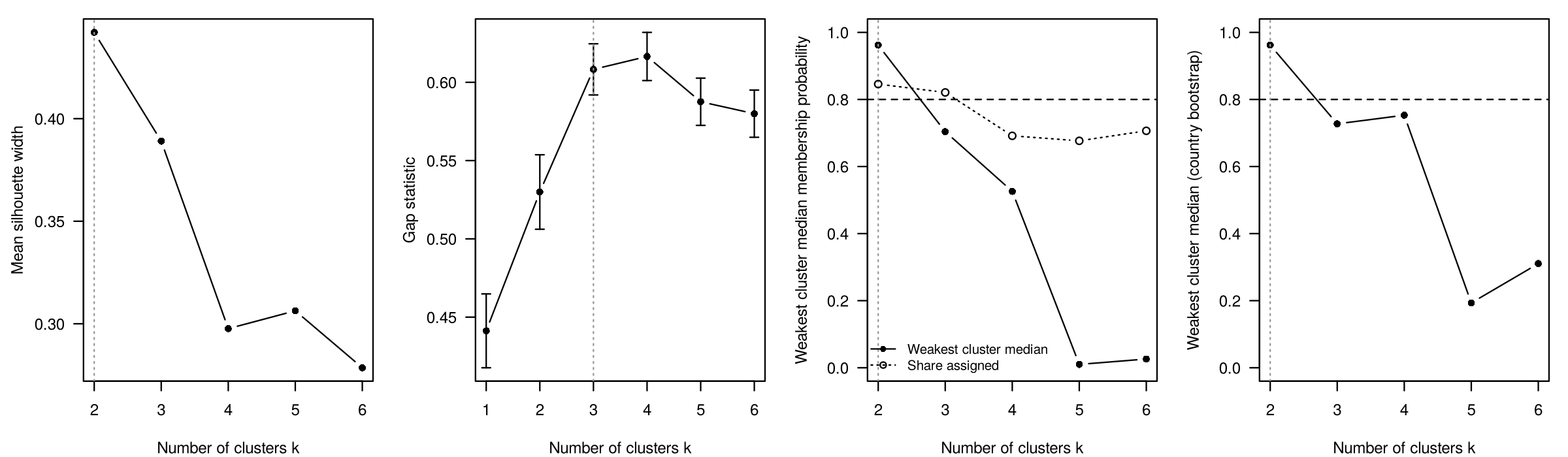}
\caption{Selection of the number of clusters. From left: mean silhouette width; gap statistic with one-standard-error bars; weakest cluster's median membership probability under value perturbation (filled) with the share of countries assigned (open); the same under country bootstrap. Dashed line, the 0.80 rule; dotted vertical, the $k$ each criterion selects.}
\label{fig:k}
\end{figure}

The three criteria disagree, and the disagreement is the paper's central exhibit (Table~\ref{tab:k}, Figure~\ref{fig:k}). Silhouette peaks at $k = 2$ (0.44). The gap statistic selects $k = 3$, with $k = 4$ marginally higher. The uncertainty aware rule selects $k = 2$: at $k = 3$ the weakest cluster's median membership probability is 0.70, below the 0.80 rule, and by $k = 5$ it is 0.01. The share of countries left unassigned rises from 15\% at $k = 2$ to 31\% at $k = 4$. The third cluster that the gap statistic recommends is made largely of countries whose published intervals are wide enough to move them into one of the other two.

The country bootstrap reaches the same conclusion by a different route (Table~\ref{tab:k}, last column). Both schemes give a weakest median of 0.96 at $k = 2$; at $k = 3$ the bootstrap gives 0.73 against 0.70 for value perturbation, and both reject it. They part company only at $k = 4$, where the bootstrap median of 0.75 is nearly a stable solution while value perturbation has fallen to 0.53: resampling countries is blind to a fourth cluster that is geometrically distinct but made of poorly-measured series. Country level membership probabilities under the two schemes are strongly correlated (Spearman 0.94 at $k = 3$, Figure~\ref{fig:robust}c), with the value scheme the more conservative for a minority of countries.

\begin{table}[t]
\centering\small
\caption{Selection of the number of clusters under three criteria. Gap statistic with standard error from 100 reference sets; membership probabilities from 1,000 draws. The uncertainty aware rule selects the largest $k$ whose weakest cluster has median membership of at least 0.80.}
\label{tab:k}
\adjustbox{max width=\textwidth}{\begin{tabular}{lrrrrr}
\toprule
k & Silhouette & Gap & Weakest cluster median p (value) & Share unassigned (value) & Weakest cluster median p (country bootstrap) \\
\midrule
2 & 0.442 & 0.530 (0.024) & 0.96 & 0.15 & 0.96 \\
3 & 0.389 & 0.608 (0.016) & 0.70 & 0.18 & 0.73 \\
4 & 0.298 & 0.617 (0.015) & 0.53 & 0.31 & 0.75 \\
5 & 0.306 & 0.588 (0.015) & 0.01 & 0.32 & 0.19 \\
6 & 0.279 & 0.580 (0.015) & 0.03 & 0.29 & 0.31 \\
\bottomrule
\end{tabular}
}
\end{table}

\subsection{The two clusters}

\begin{figure}[t]
\centering\includegraphics[width=\textwidth]{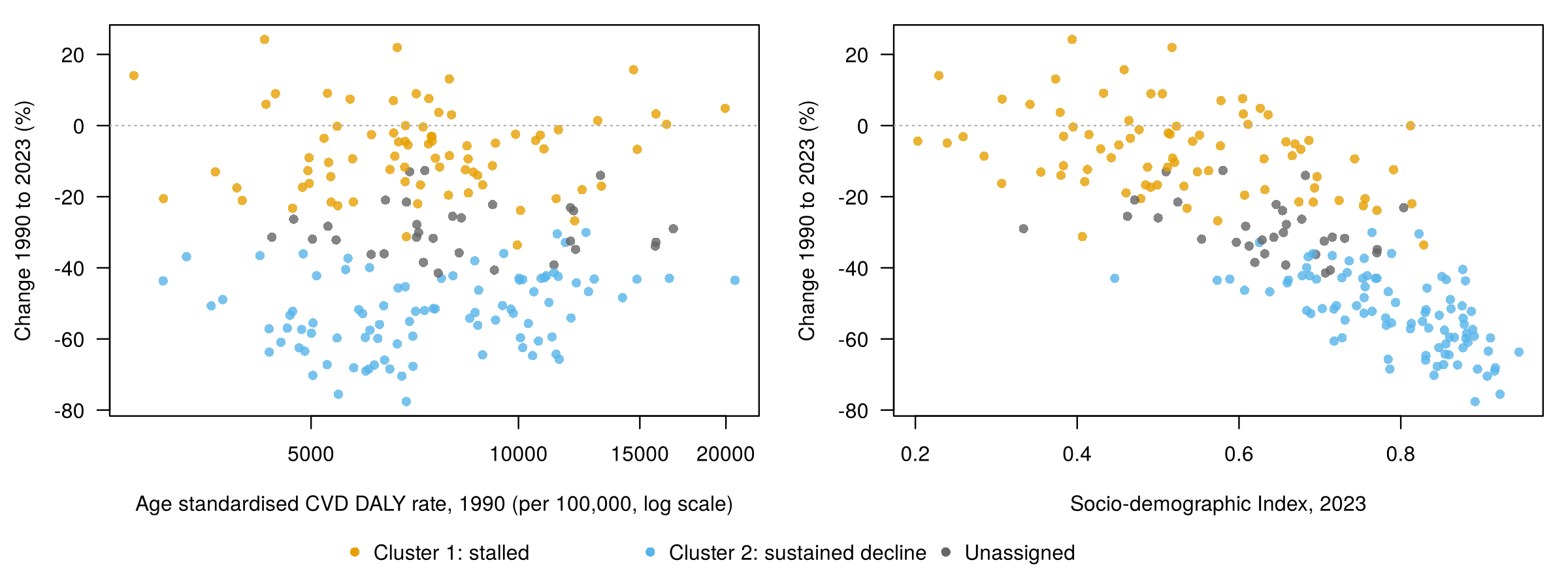}
\caption{Change in age-standardised cardiovascular DALY rate, 1990 to 2023, against the 1990 rate on a log scale (left) and against the 2023 Socio-demographic Index (right), coloured by cluster. The clusters are indistinguishable on the horizontal axis of the left panel and separate only on change.}
\label{fig:level}
\end{figure}

The decomposition into level and shape did what it was designed to do. Cluster membership explains 5.8\% of the variance in level ($R^2_L$ of~\eqref{eq:r2level}), and its adjusted Rand index against level tertiles is 0.04. The two clusters had almost identical median burden in 1990, 7,239 against 7,062 DALYs per 100,000 (Table~\ref{tab:clusters}, Figure~\ref{fig:level}a), and by 2023 stood at 6,853 against 3,589. They differ in trajectory alone.

\begin{figure}[t]
\centering\includegraphics[width=\textwidth]{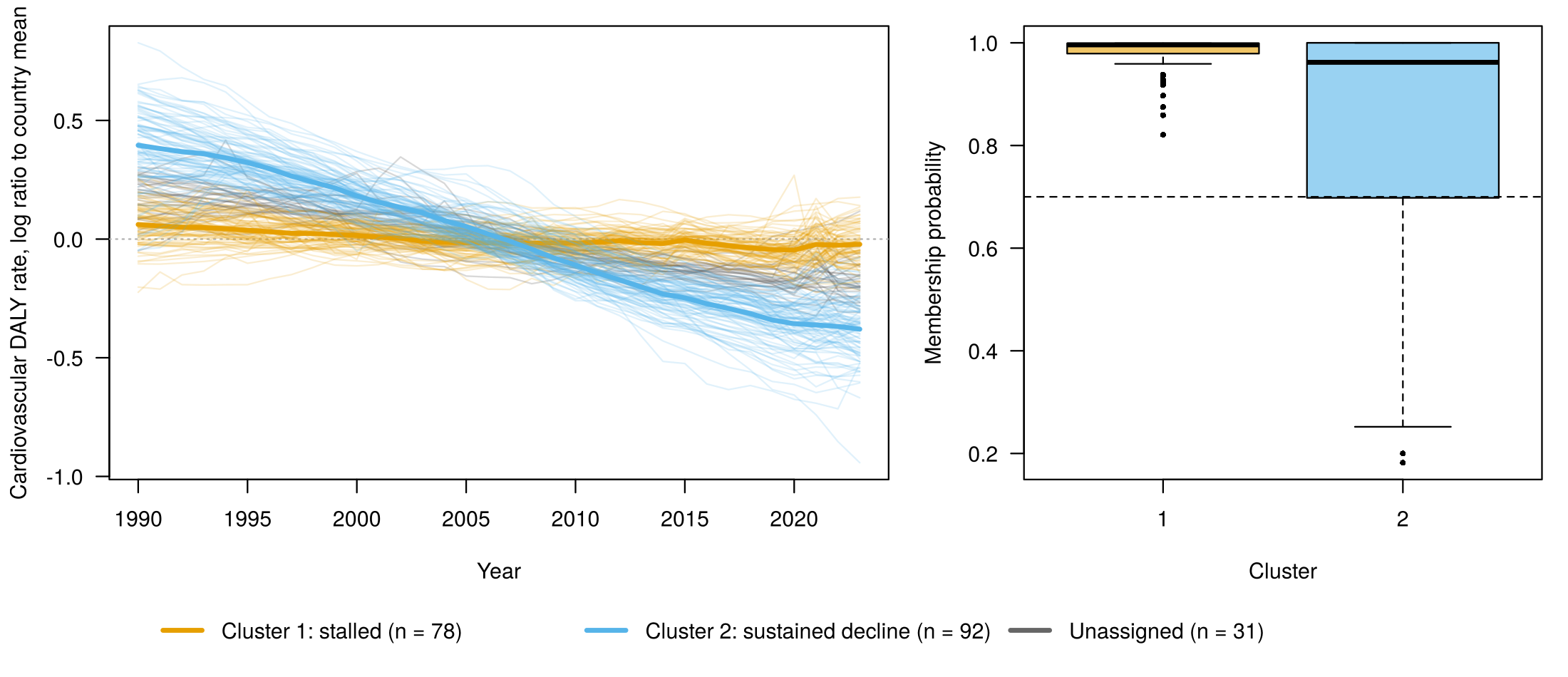}
\caption{Centred cardiovascular DALY trajectories by cluster (left) and the distribution of membership probability by cluster (right). Thin lines are individual countries, thick lines cluster means, grey lines unassigned countries. The dashed line marks the 0.70 assignment threshold.}
\label{fig:traj}
\end{figure}

\begin{figure}[t]
\centering\includegraphics[width=\textwidth]{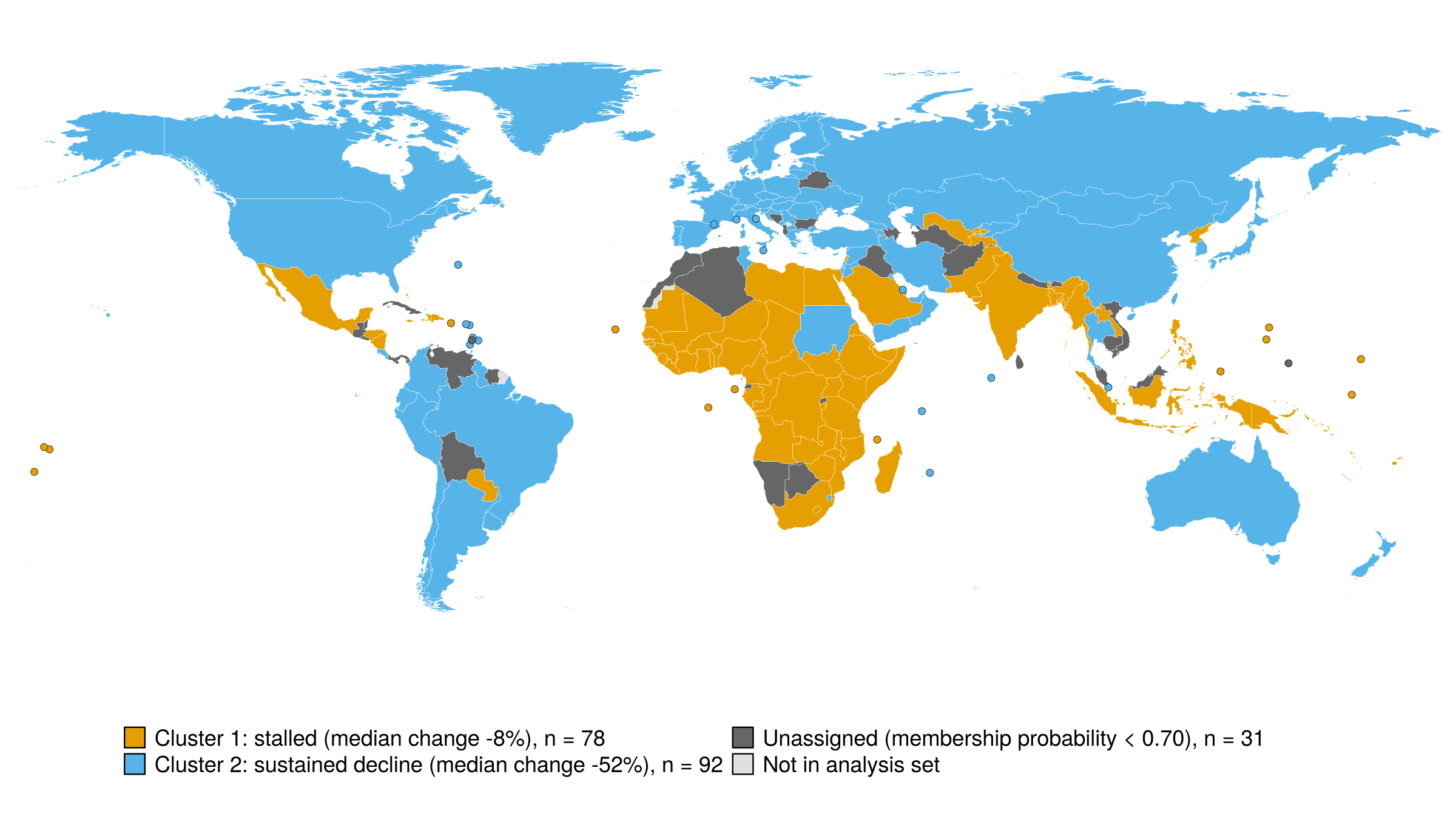}
\caption{Cluster membership for 201 countries. Orange, stalled; blue, sustained decline; dark grey, unassigned (membership probability below 0.70); light grey, not in the analysis set. Small island states are drawn as points.}
\label{fig:map}
\end{figure}

\begin{figure}[t]
\centering\includegraphics[width=\textwidth]{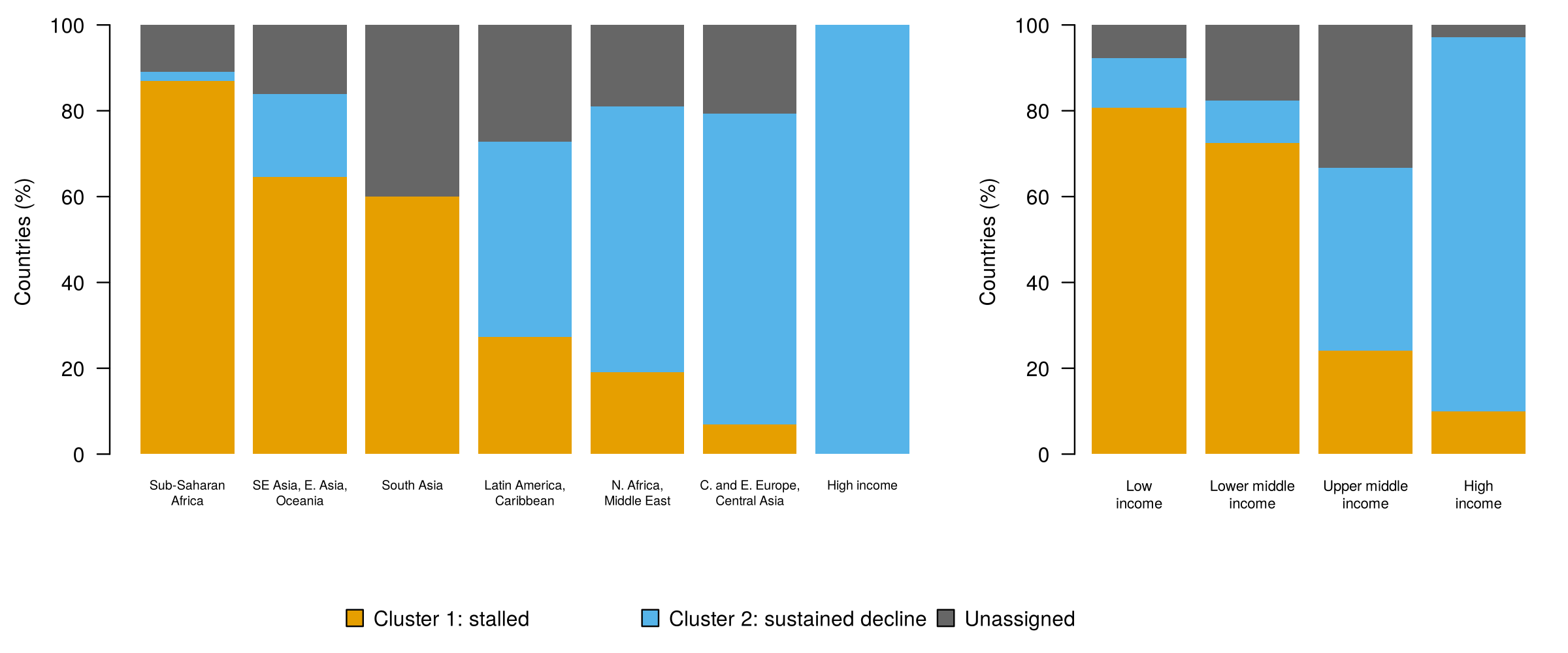}
\caption{Composition of clusters by GBD super region (left) and World Bank income group (right).}
\label{fig:comp}
\end{figure}

Cluster 1, which we label stalled, holds 78 assigned countries whose centred trajectory is nearly flat: a median change of $-7.6\%$ over 33 years, with the cardiovascular peak in the mid-1990s (median 1996, interquartile range 1990 to 2011). Forty of its members are in sub-Saharan Africa, 20 in Southeast Asia, East Asia and Oceania, and none in the high-income super region. Cluster 2, sustained decline, holds 92 assigned countries with a median change of $-52\%$ and the peak at the series start. It contains all 36 high-income super region members, 21 of the 29 countries of Central and Eastern Europe and Central Asia, China, and a minority of Latin American and Middle Eastern countries (Table~\ref{tab:region}, Figure~\ref{fig:comp}). Thirty-one countries are unassigned; they include Morocco, Algeria, Viet Nam, Nepal, Iraq, Venezuela, Botswana, Malaysia and Bulgaria, and their median change of $-31\%$ places them between the two groups. Figure~\ref{fig:traj} shows the centred trajectories and the distribution of membership probability by cluster; Figure~\ref{fig:map} maps the assignment.

\begin{table}[t]
\centering\small
\caption{Cluster characteristics, median (interquartile range). Rates are age-standardised DALYs per 100,000; change is 1990 to 2023; interval width is the country mean of the log-scale standard deviation in~\eqref{eq:sd}.}
\label{tab:clusters}
\adjustbox{max width=\textwidth}{\begin{tabular}{lrrrrrrrr}
\toprule
Group & n & Rate 1990 & Rate 2023 & Change (\%) & Peak year & SDI 2023 & Interval width & Membership probability \\
\midrule
Cluster 1: stalled & 78 & 7239 (5525 to 8836) & 6853 (5235 to 8092) & -7.6 (-16.6 to -0.3) & 1996 (1990 to 2011) & 0.52 (0.42 to 0.63) & 0.097 (0.068 to 0.109) & 1.00 (0.98 to 1.00) \\
Cluster 2: sustained decline & 92 & 7062 (5645 to 10372) & 3589 (2176 to 4877) & -52.4 (-60.0 to -43.3) & 1990 (1990 to 1991) & 0.81 (0.73 to 0.86) & 0.035 (0.027 to 0.049) & 1.00 (0.92 to 1.00) \\
Unassigned & 31 & 7520 (6645 to 11576) & 5399 (4478 to 7511) & -31.4 (-34.4 to -24.7) & 1990 (1990 to 1996) & 0.65 (0.59 to 0.70) & 0.046 (0.034 to 0.098) & 0.43 (0.34 to 0.59) \\
\bottomrule
\end{tabular}
}
\end{table}

\begin{table}[t]
\centering\small
\caption{Membership by GBD super region.}
\label{tab:region}
\adjustbox{max width=\textwidth}{\begin{tabular}{lrrrr}
\toprule
Super region & Cluster 1: stalled & Cluster 2: sustained decline & Unassigned & Total \\
\midrule
Central Europe, Eastern Europe and Central Asia &  2 & 21 &  6 &  29 \\
High income &  0 & 36 &  0 &  36 \\
Latin America and Caribbean &  9 & 15 &  9 &  33 \\
North Africa and Middle East &  4 & 13 &  4 &  21 \\
South Asia &  3 &  0 &  2 &   5 \\
Southeast Asia, East Asia and Oceania & 20 &  6 &  5 &  31 \\
Sub-Saharan Africa & 40 &  1 &  5 &  46 \\
Total & 78 & 92 & 31 & 201 \\
\bottomrule
\end{tabular}
}
\end{table}

\subsection{Attribution}

Membership is spatially coherent. Moran's I on six-nearest-neighbour weights is 0.56 for the stalled indicator (and, with two clusters, for its complement), and no permutation of 999 reached the observed value; the statistic moves only between 0.53 and 0.58 as the number of neighbours is varied from four to ten (Supplementary Table S4). Development explains about half of the partition: a logistic model of decline versus stalled on SDI alone has McFadden $R^2$ of 0.53, with the odds of the decline cluster rising 5.2-fold per 0.1 unit of SDI. Adding super region raises $R^2$ to 0.63 and the regional term is significant (likelihood ratio test, $p = 0.0005$; Table~\ref{tab:attr}). Geography adds information beyond development level, which a level-based classification would not have produced.

The interval width adds information too. Added to the SDI only model it is significant ($p = 0.011$), with the odds of the decline cluster falling by a factor of 0.11 for every 0.1 increase in mean log-scale standard deviation: at a given development level, the more a country's series depends on modelling, the more likely it is to be stalled. Added to the model that already contains region, it no longer reaches significance ($p = 0.11$), because interval width is itself strongly regional (Figure~\ref{fig:width}a): the median width is 0.024 in Central and Eastern Europe and Central Asia and 0.107 in sub-Saharan Africa. Region and measurement quality are not separable in these data, and Section~\ref{sec:discussion} returns to what that implies.

\begin{table}[t]
\centering\small
\caption{Attribution of membership (sustained decline versus stalled, assigned countries only). Likelihood ratio tests: SDI versus intercept, $p < 0.0001$; region added to SDI, $p = 0.0005$; interval width added to SDI, $p = 0.011$; interval width added to SDI and region, $p = 0.11$.}
\label{tab:attr}
\adjustbox{max width=\textwidth}{\begin{tabular}{lrrrr}
\toprule
Model & Deviance & df & AIC & McFadden R2 \\
\midrule
Intercept only & 234.5 & 169 & 236.5 & 0.000 \\
SDI & 111.0 & 168 & 115.0 & 0.527 \\
SDI + super region &  87.1 & 162 & 103.1 & 0.629 \\
SDI + interval width & 104.5 & 167 & 110.5 & 0.554 \\
SDI + super region + interval width &  84.5 & 161 & 102.5 & 0.640 \\
\bottomrule
\end{tabular}
}
\end{table}

\begin{figure}[t]
\centering\includegraphics[width=0.7\textwidth]{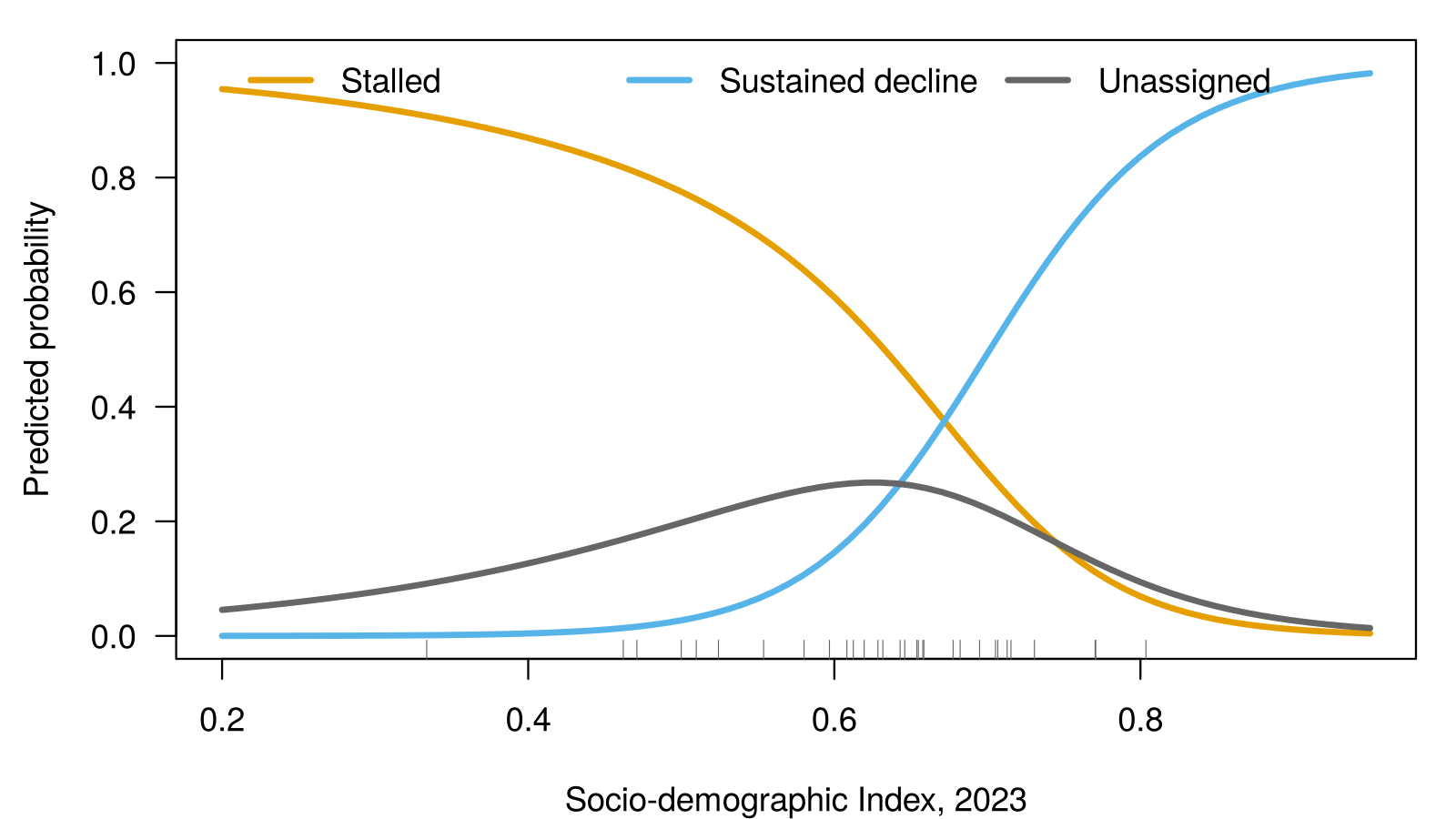}
\caption{Predicted probability of each membership outcome across the Socio-demographic Index from the multinomial model on SDI alone. Ticks on the horizontal axis mark the SDI of the 31 unassigned countries.}
\label{fig:multinom}
\end{figure}

The multinomial model over all 201 countries confirms the reading of the unassigned class as countries in transition rather than noise (Figure~\ref{fig:multinom}). SDI alone gives McFadden $R^2$ of 0.36 and super region adds significantly ($p = 0.002$). The predicted probability of being unassigned rises from near zero at low SDI to a peak of 0.26 at SDI 0.6, exactly where the stalled and decline curves cross, and falls again above 0.8. The framework declines to place countries at the point where the two-shapes are least distinguishable, which is what an honest classifier should do.

\subsection{Robustness}

\begin{figure}[t]
\centering\includegraphics[width=\textwidth]{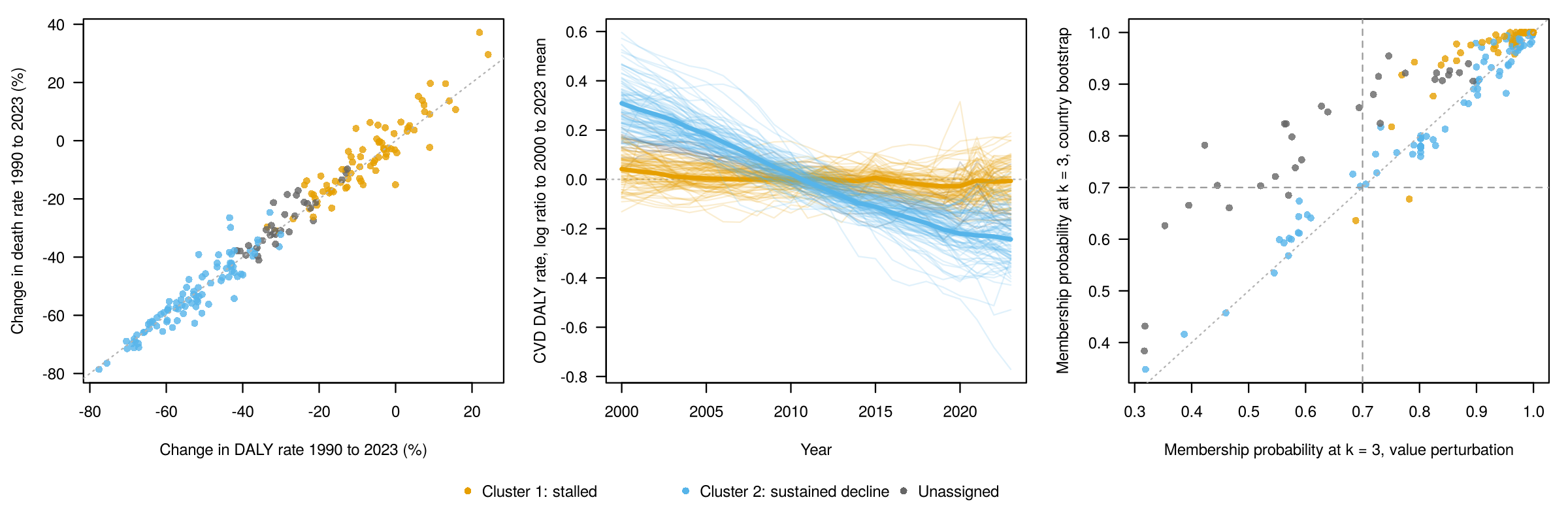}
\caption{Robustness. (a) Change in death rate against change in DALY rate, 1990 to 2023, coloured by the primary cluster. (b) Centred DALY trajectories over 2000 to 2023, coloured by the window analysis cluster. (c) Membership probability at $k = 3$ under country bootstrap against value perturbation, coloured by the primary cluster; dashed lines at 0.70.}
\label{fig:robust}
\end{figure}

The partition survives every check that the data allow (Table~\ref{tab:robust}, Figure~\ref{fig:robust}). On age-standardised death rates in place of DALYs, all three criteria return the same $k$ as on DALYs, 165 countries are assigned, and among the 156 assigned in both analyses the adjusted Rand index is 1.00: not one country moves between the stalled and decline clusters. The nine countries assigned on DALYs but not on deaths, and the nine assigned on deaths but not on DALYs, are all near the threshold in one measure and all retain the same cluster label. The stalled trajectory is not an artefact of combining mortality and disability.

On the 2000 to 2023 window, all three criteria select two clusters and 188 countries are assigned, the larger number reflecting the smaller spread of trajectory shapes over a shorter window. All 36 high-income countries remain in the decline cluster. The slowdown is real but not a cluster: the median annual change in the high-income super region fell from $-3.5\%$ over 2000 to 2010 to $-1.7\%$ over 2013 to 2023, visible as a flattening of the blue mean in Figure~\ref{fig:robust}b, yet it did not separate those countries from the rest of the decline group. The late-stalling hypothesis of a third trajectory type among wealthy countries is not supported by GBD 2023 at the resolution of this analysis. Of the 31 countries unassigned over the full window, 22 are assigned over the shorter one, ten to stalled and twelve to decline; among countries assigned in both windows the adjusted Rand index is 0.98.

\begin{table}[t]
\centering\small
\caption{Robustness of the primary partition. Agreement is the adjusted Rand index with the primary assignment over countries assigned in both analyses. The narrow half analysis uses the 101 countries with mean interval width below the median.}
\label{tab:robust}
\adjustbox{max width=\textwidth}{\begin{tabular}{lrrrrrr}
\toprule
Analysis & k (silhouette) & k (gap) & k (stability) & Weakest median p at k = 3 & Assigned & ARI with primary (jointly assigned) \\
\midrule
DALYs 1990 to 2023 (primary) & 2 & 3 & 2 & 0.70 & 170 & 1.000 \\
Deaths 1990 to 2023 & 2 & 3 & 2 & 0.66 & 165 & 1.000 \\
DALYs 2000 to 2023 & 2 & 2 & 2 & 0.44 & 188 & 0.976 \\
DALYs, narrowest half of intervals (n = 101) & 2 & 1 & 3 & 0.85 &  72 & 0.414 \\
Country bootstrap instead of value perturbation & 2 & 3 & 2 & 0.73 & 177 & 1.000 \\
\bottomrule
\end{tabular}
}
\end{table}

Varying the two tuning constants changes counts, not conclusions. The stability rule $\pi$ selects $k = 2$ for every value from 0.75 to 0.90; at $\pi = 0.70$ it admits $k = 3$, because the weakest cluster at $k = 3$ sits exactly on that boundary (median 0.70), which is why we set the rule above the assignment threshold rather than equal to it. The assignment threshold $\tau$ moves the number of unassigned countries from 24 at $\tau = 0.60$ to 53 at $\tau = 0.90$ without changing the label of any assigned country, since labels come from the reference partition and $\tau$ only gates reporting. We recommend reporting the full membership distribution so that readers can apply their own threshold.

\subsection{What the best-measured countries show}

\begin{figure}[t]
\centering\includegraphics[width=\textwidth]{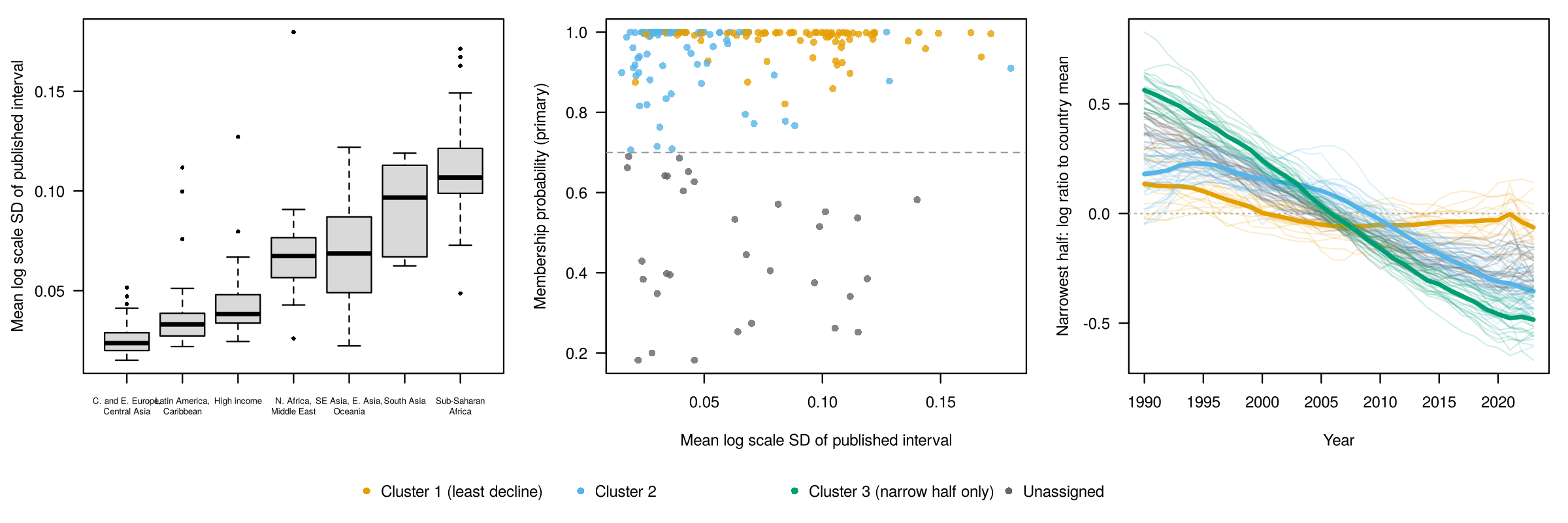}
\caption{Measurement quality. (a) Mean log-scale standard deviation of the published interval by super region. (b) Interval width against membership probability in the primary analysis. (c) Centred trajectories for the 101 countries with the narrowest intervals, coloured by the three-cluster solution found on that subset; green is the third cluster that appears only here.}
\label{fig:width}
\end{figure}

Restricting the analysis to the 101 countries with mean interval width below the median changes the picture in an instructive way (Figure~\ref{fig:width}c, Table~\ref{tab:robust}). The subset is not representative: it holds 28 of the 29 countries of Central and Eastern Europe and Central Asia, 30 of 36 high-income countries, 29 of 33 in Latin America and the Caribbean, and one country in sub-Saharan Africa. On this subset the uncertainty aware rule selects three clusters, each with median membership above 0.85, where the gap statistic finds no structure at all. The three shapes are the former Soviet and Eastern European rise and fall, with 22 of its 29 members from that super region; the high-income steady decline, with 20 of its 26 from that super region; and a stalled shape of 16 countries, eleven of them in Latin America and the Caribbean, including Mexico, Guatemala, El Salvador, Nicaragua, Paraguay, Venezuela and the Philippines.

Two conclusions follow. The stalled trajectory exists among well-measured countries, so it is not purely a modelling artefact, though among well-measured countries it is a Latin American and Pacific phenomenon rather than an African one. And the pooled analysis loses structure: the Eastern European hump, which the global solution absorbs into the decline cluster, is a distinct and stable shape once the wide-interval series that dominate the pooled variance are removed. Agreement between the subset solution and the primary partition restricted to the same countries is only 0.41 by adjusted Rand index, because the subset splits the primary decline cluster in two.

\section{Discussion}\label{sec:discussion} 

\subsection{What the framework changes} 

This study addresses a simple but largely overlooked problem: secondary analyses of modelled estimates routinely ignore the uncertainty that the original studies work hard to quantify. In the present case, a standard clustering criterion selected three groups, whereas propagating the published uncertainty through the clustering process showed that only two were stable. The difference is not technical. One analysis concludes that three cardiovascular trajectory types exist; the other concludes that only two are supported and that 31 countries cannot be assigned between them with confidence. We argue that the latter conclusion is preferable because it reflects the information actually contained in the estimates. 

More broadly, the framework shifts the object of inference from cluster assignment to assignment certainty. Countries are no longer forced into groups simply because a clustering algorithm requires a partition. Instead, membership becomes probabilistic and ambiguity becomes a result rather than a nuisance. This is particularly important for modelled estimates, whose uncertainty varies substantially across countries and over time. The approach therefore aligns the interpretation of clustering outputs with the principles underlying GATHER reporting standards \citep{stevens2016} and responds directly to concerns that uncertainty is often lost in secondary analyses of GBD estimates \citep{husoy2025}. 

The decomposition of trajectories into level and shape was also important. The two stable clusters began the study period with nearly identical cardiovascular burden yet ended more than twofold apart. The distinction between them therefore lies not in burden itself but in how burden evolved over time. Separating level from shape avoids conflating these concepts and yields more interpretable explanations of cluster membership. The resulting association with SDI and region is consequently a statement about differences in epidemiological trajectories rather than differences in absolute burden. 

The comparison between value perturbation and country bootstrap resampling provides an additional methodological insight. The two approaches agreed strongly in the present application, identifying the same number of stable clusters and yielding highly correlated membership probabilities. However, they diverged when considering more complex partitions. A cluster composed largely of poorly measured countries survived country resampling but disappeared when uncertainty in the estimated values was propagated. Unit-level stability and estimate-level stability therefore assess different forms of robustness. For analyses based on modelled estimates, both are informative, but only the latter directly evaluates the consequences of uncertainty in the estimates themselves. 

\subsection{What the cardiovascular result does and does not show} 

The substantive result is a global partition into a stalled-improvement trajectory and a sustained-decline trajectory. Broadly, this pattern is consistent with established descriptions of the cardiovascular transition \citep{gbdcvd2025,gbd2023cod}: prolonged declines in countries where prevention, treatment and risk-factor control matured early, and weaker improvements in many low- and middle-income settings. Replication using cardiovascular mortality rather than DALYs indicates that the finding is not driven by disability estimates alone. 

The analysis also clarifies what clustering cannot show. The well-documented slowdown in cardiovascular mortality decline across many high-income countries \citep{lopez2019,dowd2025,abrams2025} did not produce a separate trajectory type because the slowdown was broadly shared within that group. Trajectory clustering identifies divergence between countries, not changes common to all members of a cluster. Detecting such shifts would require methods designed to identify changes in the cluster mean rather than changes in cluster membership. 

The strongest substantive caution concerns measurement quality. Countries in the stalled cluster generally had much wider uncertainty intervals than countries in the decline cluster, and interval width predicted cluster membership beyond development level alone. This result does not imply that the stalled trajectory is an artefact of modelling, nor does it imply that it is entirely real. Rather, it shows that the observed pattern is partly entangled with differences in measurement quality. Analyses based solely on published estimates cannot fully disentangle those explanations. What the proposed framework offers is transparency: uncertainty is carried through the analysis, countries with unstable memberships are identified explicitly, and dependence on measurement quality becomes visible rather than hidden. 

At the same time, the restricted analysis of well-measured countries indicates that richer trajectory structure may exist than is apparent in the full dataset. In particular, a third trajectory emerged among former Soviet and Eastern European countries once countries with the widest uncertainty intervals were removed. This suggests that pooling countries of widely differing measurement quality can mask meaningful heterogeneity. 

\subsection{Limitations} 

Several limitations should be noted. Uncertainty propagation relied on independent log-scale perturbations derived from published uncertainty intervals and did not incorporate the temporal correlation present in the full posterior draws. Both simplifications were dictated by data access constraints and should be replaced by draw-level analyses when those draws are available. Interval width was used as a proxy for measurement quality; although informative, it captures all sources of uncertainty and is not equivalent to direct measures of data completeness such as GBD's vital registration quality ratings. 

The case study was intentionally univariate. Consequently, the identified clusters should be interpreted as cardiovascular burden trajectory types rather than comprehensive epidemiological transition types. Extending the framework to multivariate trajectories is straightforward and represents a natural next step. The explanatory models were likewise limited to development level and region; future work could incorporate health-system characteristics, behavioural risk factors and spatial dependence structures. Finally, the stability thresholds used to define supported clusters and uncertain memberships are conventions rather than theoretically determined quantities, which is why the full distribution of membership probabilities is reported. 

\subsection{Implications} 

The practical implications are straightforward. Whenever modelled estimates are clustered, uncertainty should be propagated through the clustering process rather than ignored. The number of clusters should be evaluated using stability under published uncertainty alongside conventional geometric criteria. Membership probabilities should be reported for individual units, and units with consistently ambiguous membership should remain unclassified. Measures of data quality should also be examined explicitly, particularly when cluster membership is associated with uncertainty. 

For cardiovascular disease, the principal conclusion is that countries are more reliably described by two broad trajectory types than three, that a substantial set of countries falls between those types, and that measurement quality is intimately connected to the resulting classification. More generally, the study demonstrates that uncertainty is not merely a property of the estimates used for clustering. It is a property of the clustering solution itself.

\section*{Declarations}

\paragraph{Data availability.} All inputs are public GBD 2023 outputs of the Institute for Health Metrics and Evaluation. The extracted panels, lookup tables and R scripts that reproduce every table and figure are provided as supplementary material.

\paragraph{AI assistance.} Claude (Anthropic) was used for data extraction from the GBD 2023 connector, supported the drafting and executing the analysis code, and literature search. The author reviewed and revised all outputs and takes responsibility for the content. 

\paragraph{Funding.} None declared. \textbf{Competing interests.} None declared.

\section*{Supplementary material}
Supplementary Table S2 lists all 201 countries with region, income group, SDI, change in DALYs and deaths, peak year, interval width, and cluster and membership probability under the primary, deaths and 2000 window analyses. Table S3 lists the unassigned countries, Table S4 gives Moran's I for four to ten nearest neighbours, Table S5 the sensitivity to the tuning constants, Table S6 the selection diagnostics for every robustness run, and Table S7 the composition of the narrow-interval solution. Section S1 documents the extraction and Section S6 the reproduction steps. The R scripts \texttt{final\_analysis.R} and \texttt{extended\_analysis.R}, the four panels and the two lookup tables reproduce every result. Codes and extracted data are available at \url{https://github.com/EAfrifa-Yamoah/Clustering_trajectories_of_modelled_estimates}.

\bibliographystyle{plainnat}
\bibliography{refs}

\end{document}